\documentclass[aps,prb,groupedaddress,twocolumn]{revtex4}
\usepackage{epsfig}
\usepackage[dvipsnames,usenames]{color}
\usepackage{color}
\usepackage{amsmath}
\usepackage{comment}
\usepackage{array}
\usepackage{multirow}
\usepackage{tabularx}
\usepackage{float}
\usepackage{hyperref}

\emergencystretch=\maxdimen
\begin{document}
\title{Antiferromagnetism mediated by heavy electrons: \\ singlet shielding vs ``high-order'' RKKY}
\author{Mi Jiang}
\affiliation{Stewart Blusson Quantum Matter Institute, University of British Columbia, Vancouver, BC, Canada}

\begin{abstract}
Can the antiferromagnetic (AF) order be induced via the local moments' hybridization with the heavy electrons instead of conduction electrons? We address this intriguingly fundamental question via a prototypical model to describe the interplay between local moments and heavy electrons.
We discover that the AF order can be mediated via the heavy electrons through the mechanism of ``high-order'' Ruderman-Kittel-Kasuya-Yosida (RKKY) interaction. Moreover, the induced AF order can coexist with its metallicity in a finite regime of the phase diagram, which competes with and ultimately destroys the AF order concurrently with the breakdown of heavy electrons. The potential relevance to the heavy fermion compound Ce$_3$(Pt/Pd)In$_{11}$ is discussed.
\end{abstract}

\maketitle

\section{Introduction}
The conventional Kondo/Anderson lattice model (KLM/PAM) describes the competition of antiferromagnetism and Kondo screening as a fundamental model of heavy fermion physics~\cite{pam1,pam2,pam3,pam4,pam5,PAMreference}. 
Generally, there are two well-known exchange mechanism responsible for the formation of the antiferromagnetic phase, i.e. (1) the superexchange interaction between localized electrons via their hybridization with the conduction band and (2) the Ruderman-Kittel-Kasuya-Yosida (RKKY) interaction originating from the scattering of the conduction electrons from two localized moments~\cite{Freericks1997}. The former always favors the antiferromagnetic order between localized moment but is strongly suppressed if the conduction electron band approaches to half filled~\cite{superexchange}; while the latter's sign and magnitude vary with the distance between localized moments and also the conduction band filling~\cite{Fye1990}.
Take the conventional PAM as an example, it is well known that at small hybridization between the conduction and localized electrons, the indirect RKKY interaction induces the antiferromagnetic ground state, which competes with the paramagnetic spin liquid ground state formed by Kondo screening the local electrons by the conduction band at large hybridization.

The common thread between the aforementioned two exchange mechanism relies on the conduction electrons. It is natural to ask whether or not the heavy electrons can similarly act as a ``glue'' for the antiferromagnetic (AF) order between local moments. Until now, surprisingly, there has been few studies on this fundamental question despite that multiple 4f orbitals was considered in the context of Cerium volume collapse considering the inherent 4f electronic correlations~\cite{CeVolCollapse}. We point out that the AF order mediated by heavy electrons is not only an abstract theoretical question but also relevant to recent discovery of the microscopic coexistence between AF and superconductivity in a particular family of heavy fermion compounds Ce$_3$(Pt/Pd)In$_{11}$ harboring two inequivalent Ce sites~\cite{Custers2015,Custers15,Kaczorowski18,Custers2020}, where the most fascinating scenario proposed for their coexistence claims that the Ce(1) sublattice is fully Kondo screened and responsible to the superconducting state while the Ce(2) sublattice forms the magnetic ordering. Therefore, one intrinsic problem is the interplay between Ce(1) and Ce(2) sublattices, particularly the possible Ce(2) magnetic order induced by Ce(1) sublattice.

Motivated by these experimental progress, in this proof of principle study, we explore the possibility of AF order mediated by heavy electrons via a prototypical model to describe the interplay between the local moments and heavy electrons.
Specifically, we discover that the hybridization between the local moments and heavy electrons can indeed induce AF order between the local moments through the so-called ``high-order'' RKKY interaction that resembles the conventional one mediated via the conduction electrons in standard PAM/KLM.


\section{Model and methodology}
To illustrate our findings, we adopt the simplest and prototypical model consisting of two distinct localized $f$-orbitals together with the conduction electrons on two-dimensional square lattice, which reads in the half-filled form:
\begin{eqnarray}
    {\cal H} &=& - t \sum\limits_{\langle ij \rangle \sigma}
(c^{\dagger}_{i\sigma}c_{j\sigma}^{\vphantom{dagger}}
+c^{\dagger}_{j\sigma}c_{i\sigma}^{\vphantom{dagger}}) 
- \mu \sum\limits_{i\sigma} (n^{c}_{i\sigma}+ n^{f_1}_{i\sigma} +n^{f_2}_{i\sigma})  \nonumber \\
&+& V \sum\limits_{i \sigma}  (c^{\dagger}_{i\sigma}f_{1i\sigma}^{\vphantom{dagger}}+ f^{\dagger}_{1i\sigma}c_{i\sigma}^{\vphantom{dagger}})
 + t_{\perp} \sum\limits_{i \sigma}  (f^{\dagger}_{1i\sigma}f_{2i\sigma}^{\vphantom{dagger}}+ f^{\dagger}_{2i\sigma}f_{1i\sigma}^{\vphantom{dagger}}) \nonumber \\
    &+& U \sum\limits_{mi} (n^{f_m}_{i\uparrow}-\frac{1}{2}) (n^{f_m}_{i\downarrow}-\frac{1}{2})
\label{model}
\end{eqnarray}
where $c^{\dagger}_{i\sigma}(c_{i\sigma}^{\vphantom{dagger}})$
and $f^{\dagger}_{mi\sigma}(f_{mi\sigma}^{\vphantom{dagger}})$ with $m=1,2$
are creation(destruction) operators for conduction and two local $f_{1,2}$ electrons on site $i$ with spin $\sigma$.
$n^{c,f_m}_{i\sigma}$ are the associated number operators. The chemical potential $\mu$ can be tuned for a desired average occupancy of three orbitals.
The hopping $t=1$ between conduction electrons on
nearest neighbor sites $\langle ij \rangle$ sets the energy scale. $U$ is the local repulsive interaction in the $f_{1,2}$ orbital. Note that in this work we only consider the case that the $f_{1,2}$ orbitals share an identical $U$ for simplicity although generally they can differ. The two remaining control parameters are two distinct hybridizations, namely $V$ between $c$-$f_1$ and $t_\perp$ between $f_1$-$f_2$.

Before proceeding, we remark that in the heavy fermion compounds with multiple crystallographic inequivalent local moment sites, such as Ce$_3$(Pt/Pd)In$_{11}$~\cite{Custers2015,Custers15,Kaczorowski18,Custers2020}, the local environment of two Ce ions are different leading to distinct Kondo interaction strengths with the conduction electrons. Here our focus is on the sole effects of the heavy electrons from $c$-$f_1$ Kondo singlets on the additional $f_2$ local moments. Thus, we neglect the $c$-$f_2$ hybridization to avoid the additional Kondo screening from $c$ electrons and associated complexity. The more realistic modelling of the heavy fermion compounds is left for future investigation. 

In addition, to explicitly investigate the AF order without charge fluctuation, we stick on the half-filled systems by setting $\mu=0$ so that the $c$ and $f_{1,2}$ orbitals are individually half-filled, which also ensures that the superexchange interaction between local moments is suppressed~\cite{superexchange} as discussed before so that only the RKKY-type interaction can take the role of mediating the AF order. 

To gather some initial insights of this model, it is worthwhile elaborating on some limiting cases.
In the absence of Hubbard interaction $U$, the three-orbital unit cell gives rise to three energy bands such that the system hosts a metallic ground state for any finite $V, t_{\perp}$ at half-filling. As discussed later, turning on $U$ opens the orbital-selective spectral gap. In the extreme case of $t_{\perp} \ll V$, the system separates into conventional PAM plus additional individual local moments; in contrary, if $t_{\perp} \gg V$, the system becomes a conduction band plus individual strongly bound dimers.

To fully take into account all the energy scales on the equal footing, we use the well established numerical technique of finite temperature determinant Quantum Monte Carlo (DQMC)~\cite{blankenbecler81} to explore the physics of Eq.~\ref{model}. As a celebrated computational method, DQMC provides an approximation-free solution in the presence of strong correlations.
Besides, finite size scaling can facilitate the extraction of the AF order parameter reliably so that all the quantities throughout the paper are extracted values in the thermodynamic limit.

Throughout the paper, we concentrate on the characteristic intermediate coupling strength $U=4.0t$, where it has been widely believed that the critical $c$-$f$ hybridization strength separating the Kondo singlet and antiferromagnetic insulating ground states in PAM is $V_c \sim 1.0t$~\cite{wenjian,pam5}. Because the major purpose of this work is the AF order induced by heavy electrons, we only explore the systems with $V/t \ge 1.2$ such that the $c$-$f_1$ subsystem is readily within the heavy electron regime. Besides, all the physical quantities are obtained through finite-size scaling in lattice sizes as large as $N=10 \times 10$ at lowest temperature $T=0.025t$ with periodic boundary.

\begin{figure}[b] 
\psfig{figure=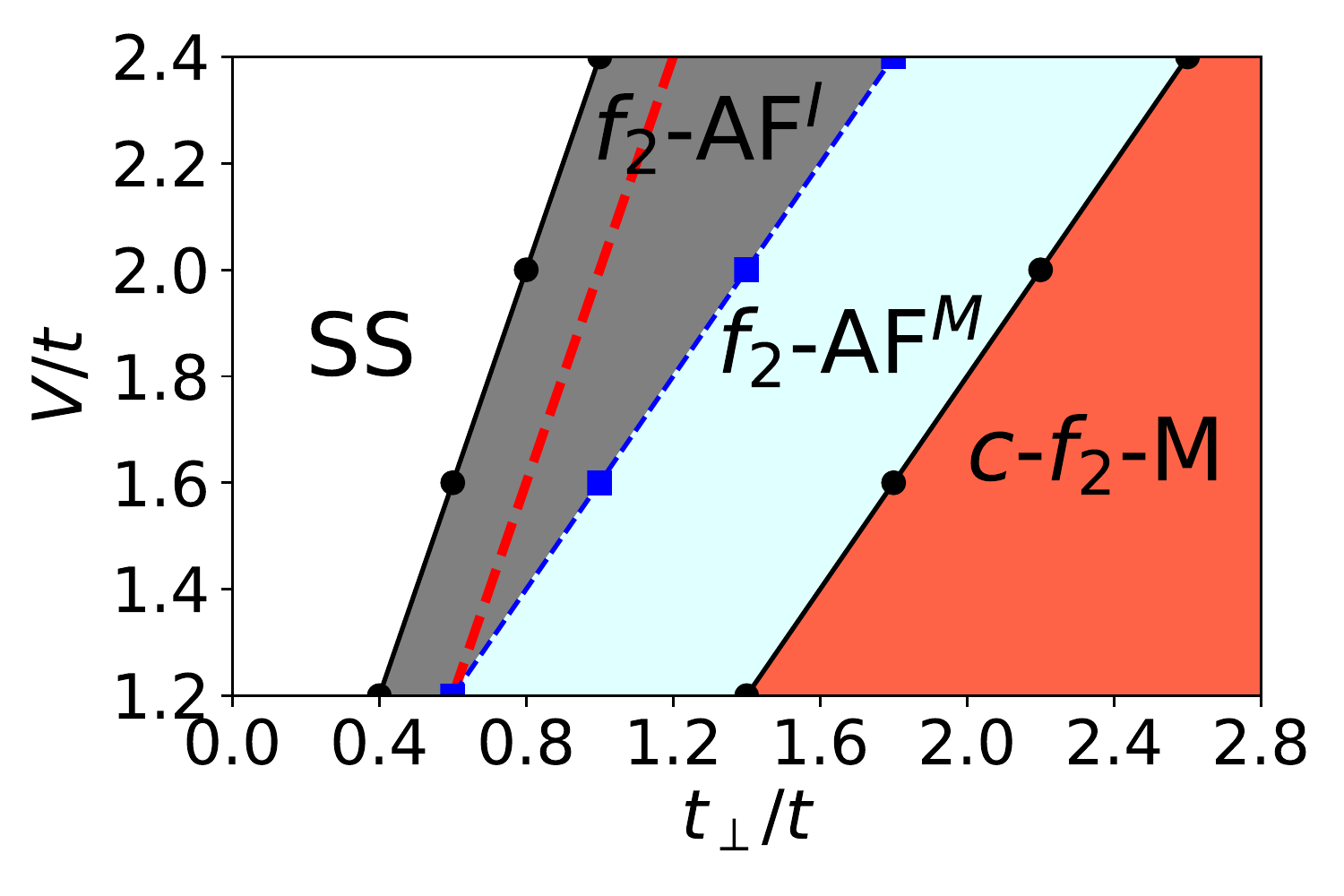}.pdf, height=6.2cm,width=.5\textwidth,angle=0,clip}
\caption{Tentative phase diagram of Eq.~\ref{model} at half-filling at lowest simulated temperature $T=0.025t$. The gray and cyan regimes exhibit the $f_2$-AF order mediated via heavy electrons ($c$-$f_1$). The red dashed line highlights the position with maximal AF structure factor. See text for details.}
\label{phase}
\end{figure} 

Our major findings are illustrated in the tentative phase diagram Figure~\ref{phase} and summarized as follows:
\begin{enumerate}
\itemsep0em 
\item {\bf Singlet shielding (SS)}: $f_2$ local moments are effectively standing alone and shielded from the heavy electrons ($c$-$f_1$ singlets);
\item {\bf $f_2$-AF$^I$}: $f_2$-AF ordered insulator via ``high-order'' RKKY coexisting with $c$-$f_1$ heavy electrons;
\item {\bf $f_2$-AF$^M$}: $f_2$-AF order with metallic feature coexisting with partially broken heavy electrons (metallic $c$ electrons);
\item {\bf Orbital-selective metallicity}: $f_2$-AF disappears due to broken heavy electrons; both $c$ and $f_2$ exhibit metallicity while $f_1$ remains insulating.
\end{enumerate}

Some remarks follow in order. 
First of all, in all phases, the $f_1$ orbital exhibits the stable insulating behavior even though $c$ and $f_2$ exhibit metallic behavior at large $t_\perp$. Besides, $f_1$ orbital does not host the AF order unless at relatively small $V/t=1.2, 1.6$ due to the combined effects of proximity effect from $f_2$-AF and weakened $c$-$f_1$ Kondo screening.
Secondly, there is a crossover (blue dashed) line separating the $f_2$-AF$^I$ and $f_2$-AF$^M$ phases, which is supported by the strong metallic tendency of $c$ and $f_2$ orbitals that competes with and ultimately destroys the $f_2$-AF order. 
Finally, at sufficiently large $t_\perp$, $c$-$f_2$-M phase will be replaced by strongly coupled $f_1$-$f_2$ dimers together with the free conduction electrons. 
In Fig.~\ref{phase}, in addition, the red dashed line highlights the $t_{\perp}$ at which the $f_2$-AF structure factor reaches its maximum. Note that the $f_2$-AF rapidly turns on upon its emergence and gradually disappears. 
In what follows, we start providing the concrete numerical evidence to support the phase diagram in detail.

\section{Magnetic properties}
\begin{figure}[h!] 
\psfig{figure=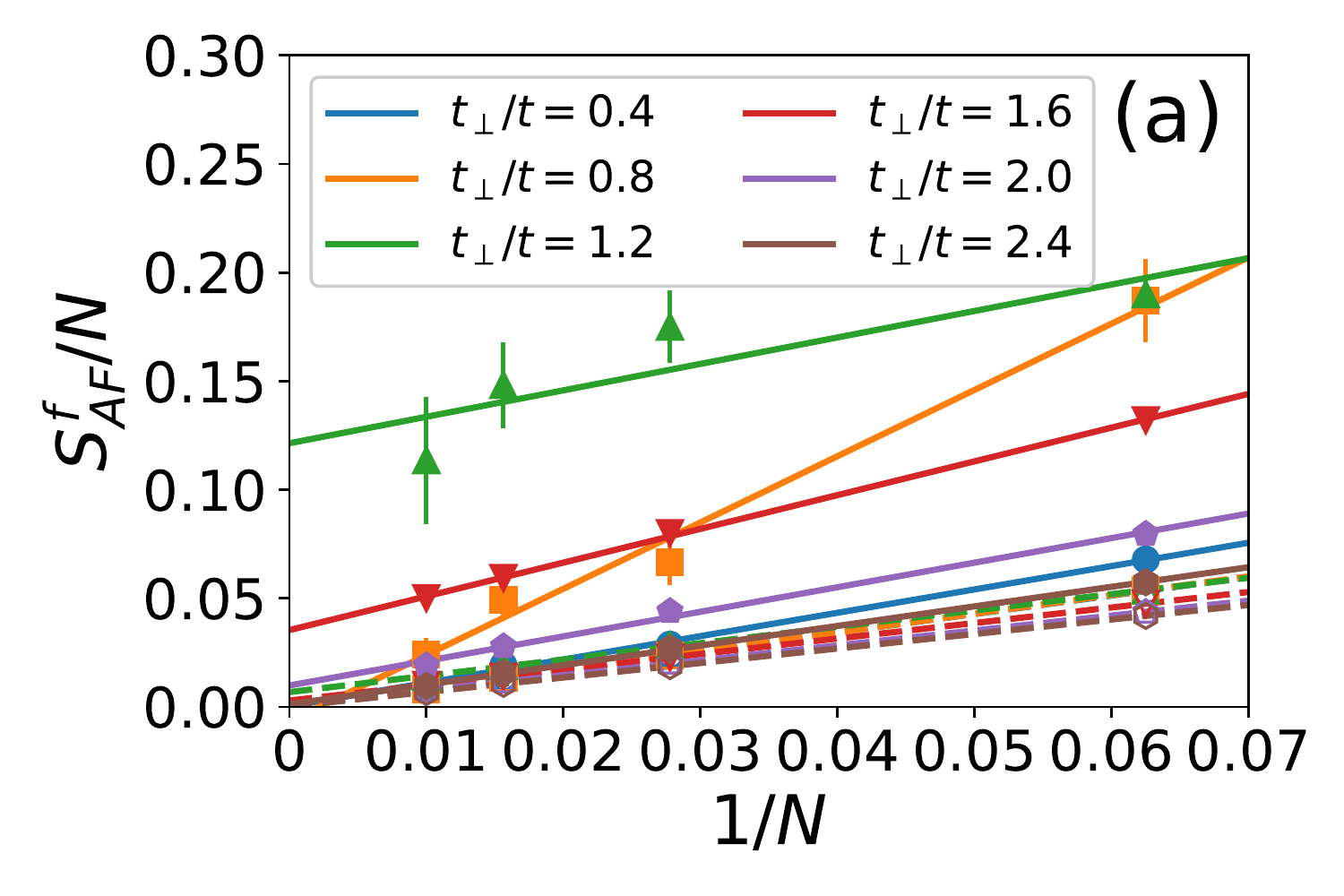}.pdf,
width=.5\textwidth,angle=0,clip} 
\psfig{figure=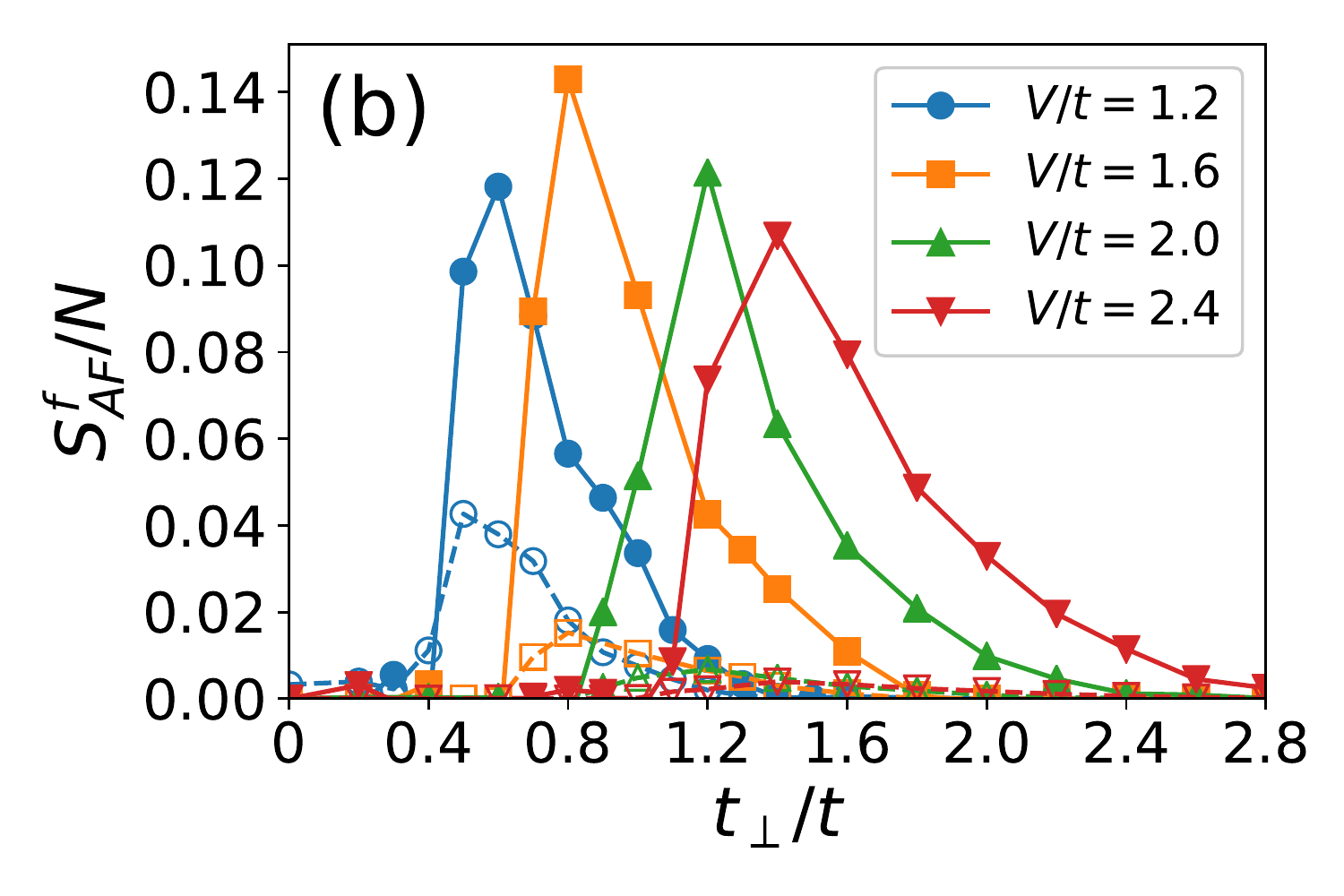}.pdf,
width=.5\textwidth,angle=0,clip=true, trim = 0.0cm 0.5cm 0.0cm 0.2cm} 
\caption{(a) Finite-size scaling of $S^{f_m}_{AF}/N$ at fixed $V/t=2.0$; (b) Evolution of extrapolated $S^{f_m}_{AF}/N$ with $t_{\perp}$ for diverse $V$ at $T=0.025t$. The dashed (solid) lines are for $f_1$ ($f_2$) orbital.}
\label{saf}
\end{figure}

We first illustrate our findings of the induced $f_2$-AF order mediated via heavy electrons ($c$-$f_1$ singlets) by the magnetic properties. 
The AF order is manifested by the AF structure factor of the $f_{1,2}$ local moments 
\begin{equation}
S^{f_m}_{AF}(V_1,V_2)=\frac{1}{N} \sum_{ij} e^{-i \mathbf{q} \cdot (\mathbf{R_i}-\mathbf{R_j})} \langle (n^{f_m}_{i\uparrow}-n^{f_m}_{i\downarrow}) (n^{f_m}_{j\uparrow}-n^{f_m}_{j\downarrow}) \rangle
\end{equation}
with $m=1,2$ at $\mathbf{q}=(\pi,\pi)$, where $\mathbf{R_i}$ denotes the coordinates of site $i$ and $N$ is the lattice size.

Figure~\ref{saf}(a) shows the finite-size scaling of $S^{f_m}_{AF}/N$ at fixed $V/t=2.0$ with dashed/solid lines denoting $m=1,2$ respectively. Clearly, at small $t_{\perp}$, the $c$-$f_1$ singlets shields the additional $f_2$ local moments and the two subsystems are effectively separated so that both show absence of AF order. With increasing $t_{\perp}$, the $f_2$-AF order emerges while $f_1$ local moments remain forming the Kondo singlets with the conduction electrons, which indicates that the $f_2$-AF order is not induced by the proximity effect from a ``$f_1$-AF'' order liberated by the $t_{\perp}$ hybridization. In the standard PAM ($t_{\perp}=0$), the RKKY interaction scales as $\sim J^{2}/W$ with $J\sim V^{2}/U$ and $W$ the conduction bandwidth. Here we claim that $f_2$-AF is realized through a mechanism of ``high-order'' RKKY interaction with modified $J\sim V^2t_{\perp}^2/U$ via an indirect $c$-$f_2$ hybridization, which competes with the Kondo screening scaling as $\sim W e^{-W/J}$. 
As expected, further increasing $t_{\perp}$ leads to the gradual diminish of $f_2$-AF due to the strong $f_1$-$f_2$ hybridization, which finally results in individual strongly bound dimers.

To make further progress, Fig.~\ref{saf}(b) demonstrates the evolution of extrapolated $S^{f_m}_{AF}/N$ with $t_{\perp}$ for diverse $V$, where the general peak structure of $f_2$-AF order (solid lines) and the absence of $f_1$-AF order in most cases (dashed lines) can be seen. Additionally, the $f_2$-AF rapidly turns on upon its emergence and gradually disappears. It is natural that stronger $V$ requires larger critical $t_{\perp}$ to overcome the $c$-$f_1$ Kondo screening to partially liberate the conduction electrons for its essential role in mediating the ``high-order'' RKKY that induces $f_2$-AF order. In contrary, only the systems of ``light'' heavy electrons with relatively small Kondo screening, e.g. $V/t=1.2, 1.6$ (blue and orange dashed lines) clearly exhibit the $f_1$-AF order whose maximum are concomitant with that of $f_2$-AF. This observation indicates the feedback among $c$-$f_1$-$f_2$ orbitals: (a) $t_{\perp}$ tends to break the heavy electrons to liberate the $c$-electrons; (b) $c$-electrons mediate the $f_2$-AF order via ``high-order'' RKKY; (c) the induced $f_2$-AF order has proximity effect to induce the potential $f_1$-AF order unfavored by heavy electrons.
Certainly, the partially liberated $c$-electrons can also mediate the $f_1$-AF order to some extent, although their combined effects quickly decay with enlarging the $c$-$f_1$ hybridization $V$.

\begin{figure}[h!] 
\psfig{figure=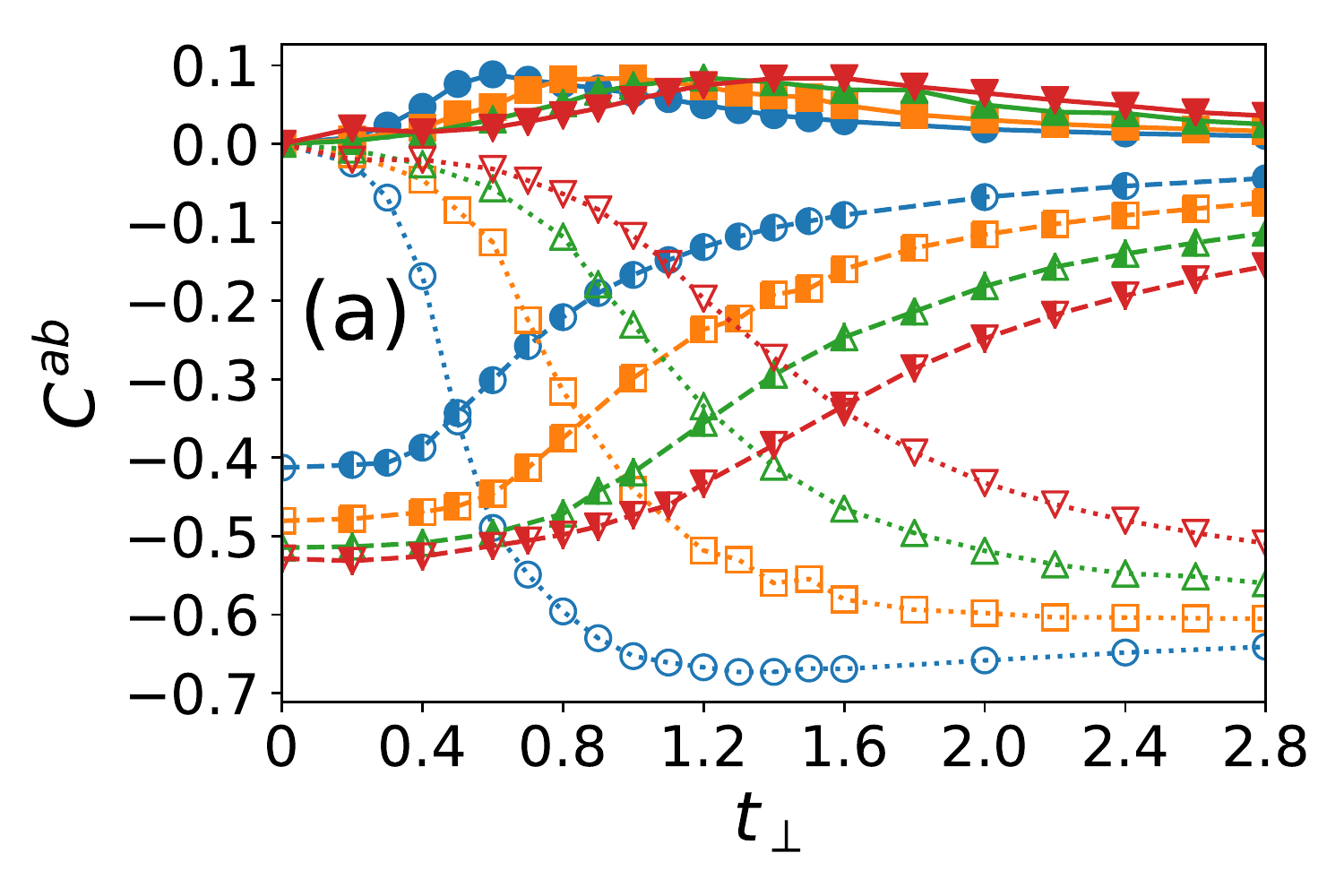}.pdf,
width=.5\textwidth,angle=0,clip=true, trim = 0.0cm 2.0cm 0.0cm 0.0cm} \\
\psfig{figure=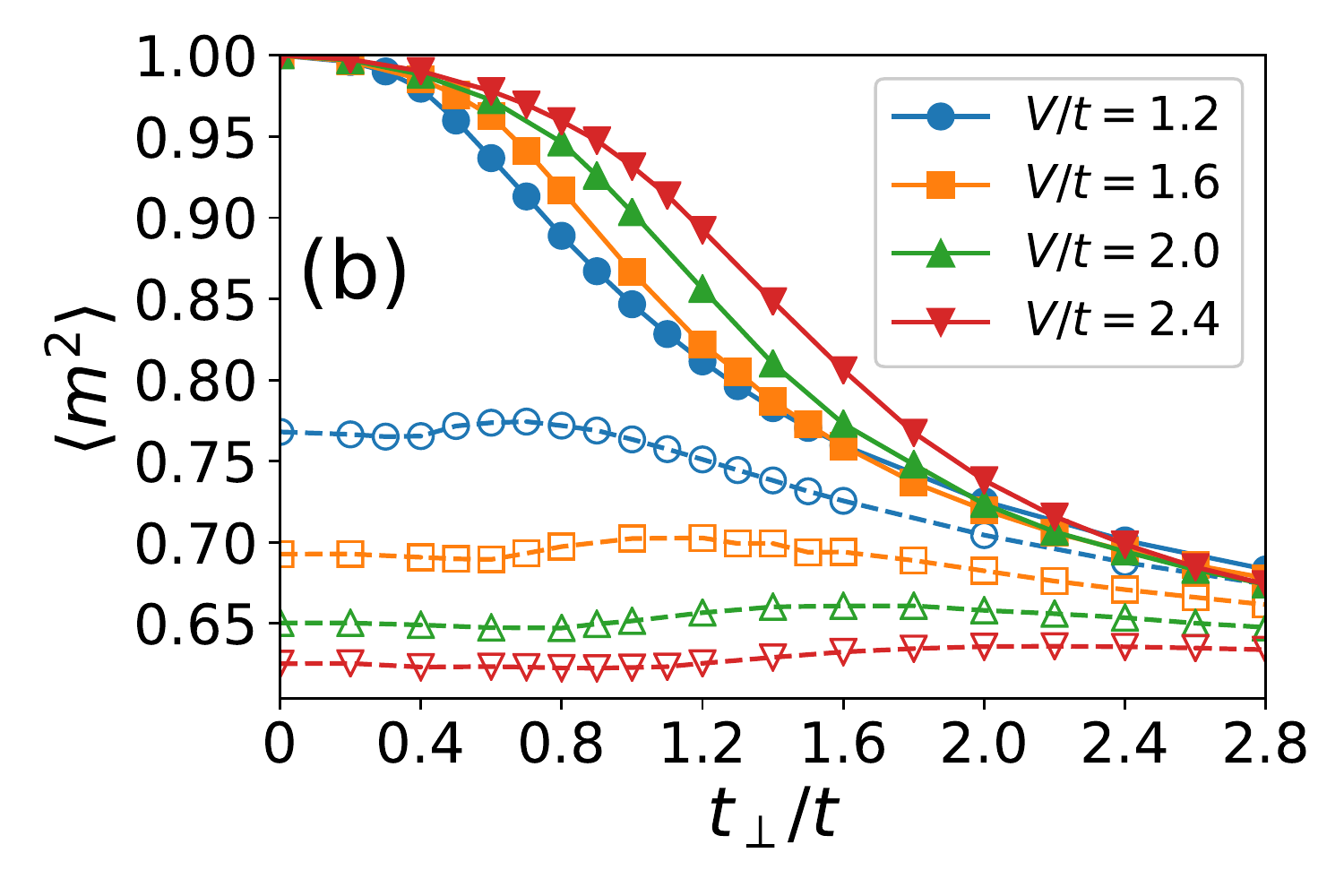}.pdf, width=.5\textwidth,angle=0,clip}
\caption{(a) Local spin correlations $C^{ab}$ between orbitals: $C^{cf_2}$ (full symbols), $C^{cf_1}$ (half-filled symbols), and $C^{f_1f_2}$ (unfilled symbols) (b) local moments $\langle m^2 \rangle$ of $f_{1}$ (dashed lines) and $f_2$ (solid lines) versus $t_{\perp}$ for diverse $V$ at $T=0.025t$.}
\label{fig3}
\end{figure}  

To further support our scenario of ``high-order'' RKKY, we resort to the local spin correlations $C^{ab}=\langle (n^a_{\uparrow}-n^a_{\downarrow}) (n^b_{\uparrow}-n^b_{\downarrow}) \rangle$ between three orbitals $a,b=c,f_1,f_2$ in Figure~\ref{fig3}(a). Apparently, $C^{cf_1}$ ($C^{f_1f_2}$) decreases (increases) in magnitude with turning on $t_{\perp}$. Nonetheless, the striking difference shows up in the indirect $C^{cf_2}$ correlation, which exhibits a nontrivial peak, whose position is consistent with the maximal $S^{f_2}_{AF}$ shown in Fig.~\ref{saf}. This strongly indicates that the ``heavy'' $c$ electrons dressed by $f_1$ local moments are mediating the $f_2$-AF order in an indirect ``high-order'' manner. More careful comparison reveals that this common peak occurs at the specific $t_{\perp}$ where $C^{cf_1} \approx C^{f_1f_2}$ and also changes most rapidly. This observation implies that the homogeneous $C^{cf_1}$ and $C^{f_1f_2}$ spin correlations are favored for enhancing $C^{cf_2}$ and in turn strengthening the ``high-order'' RKKY interaction to mediate the $f_2$-AF order. In addition, the rapid evolution of $C^{cf_1}$ and $C^{f_1f_2}$ in this regime reflects the crucial delicate balance between $C^{cf_1}$ and $C^{f_1f_2}$. Furthermore, all these observations vividly implies the vital role of the heavy electrons ($c$-$f_1$ singlets) in mediating the $f_2$-AF order. Note that $C^{cf_1}$ gradually vanishes at large $t_{\perp}$ denoting the breakdown of heavy electrons, where the $f_2$-AF order disappears concurrently.

The essential physics of our model can be also described in the viewpoint of the competition and balance between $t_{\perp}$ and $V$, which can be  explored by investigating another indicator of the magnetic properties, namely the local moments $\langle m^2 \rangle$ of $f_{1,2}$. Fig.~\ref{fig3}(b) illustrates its behavior of $f_1$ (dashed lines) and $f_2$ (solid lines). Naturally, the $f_2$ local moment decreases with $t_{\perp}$, which is most rapidly in the regime where the $f_2$-AF order emerges. Nevertheless, the $f_1$ local moment does not vary much but only possesses a bump in the regime with the maximal $f_2$-AF order, which can be traced to its quantum fluctuation subject to two-fold hybridization with $c$ and $f_2$. This provides further evidence on the steadily frozen behavior of $f_1$ orbital, whose major role is to dress the $c$ electrons.

\section{Spectral properties}
\begin{figure}[h!] 
\psfig{figure=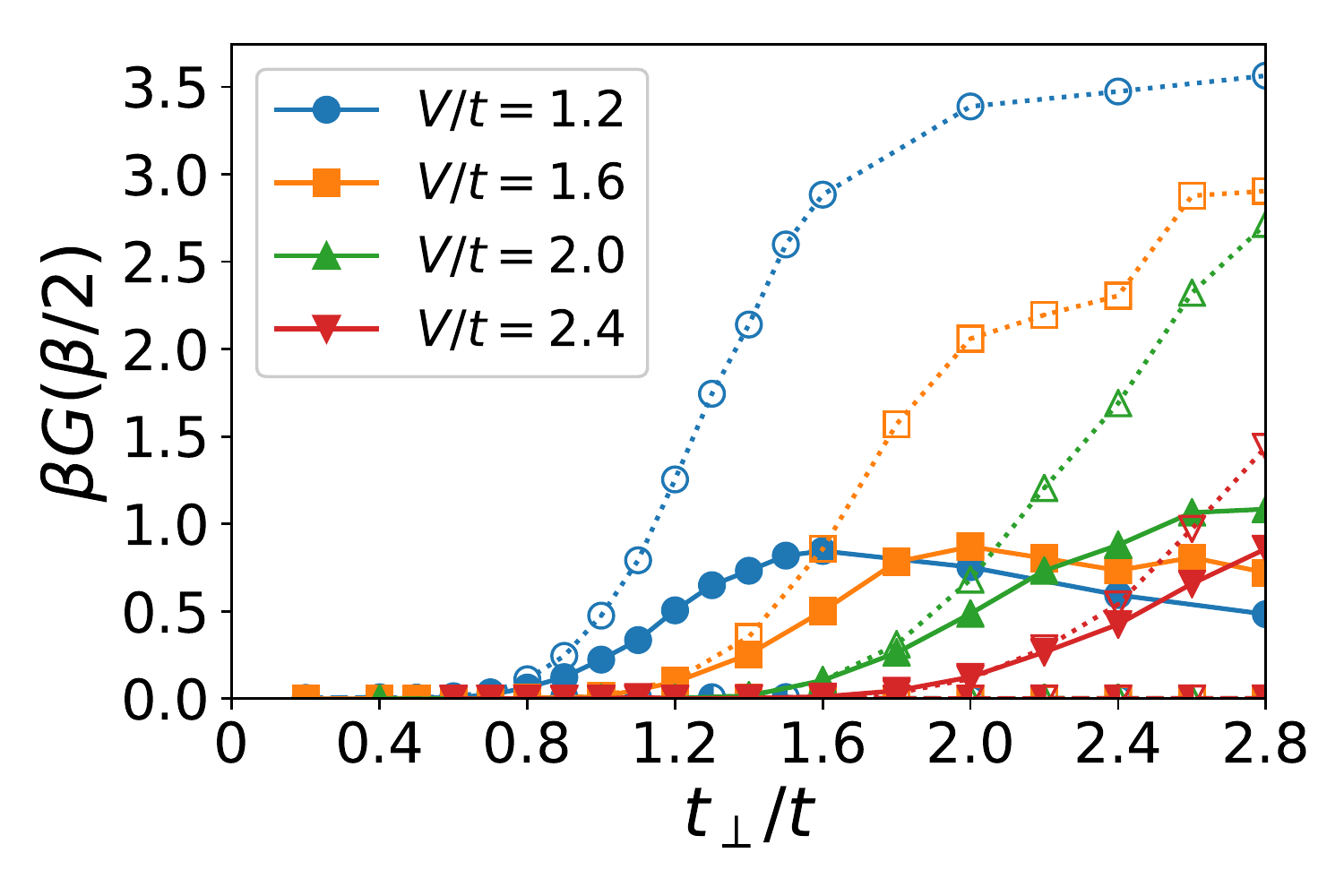}.pdf,
width=.5\textwidth,angle=0,clip=true, trim = 0.0cm 0.5cm 0.0cm 0.2cm} 
\caption{Approximate local DOS at Fermi level of three orbitals $f_2$ (full symbols), $f_1$ (half-filled symbols), and $c$ (unfilled symbols) versus $t_{\perp}$ for diverse $V$ at $T=0.025t$.}
\label{fig4}
\end{figure}  

At this stage, we have mainly focussed on the SS and $f_2$-AF$^I$ regimes at moderate $t_{\perp}$ in the phase diagram Fig.~\ref{phase}. To further understand the $f_2$-AF$^M$ and $c$-$f_2$-M regimes, we have to rely on the spectral properties. The specific question we want to address is the fate of the $f_2$-AF order at large $t_{\perp}\geq V$. To this aim, 
we examined the single-particle orbital-dependent local density of states (DOS) $N^{a}(\omega)$ with $a=c,f_1,f_2$ relating to the local imaginary-time Green's function
$G^{a}(\tau)=- \sum\limits_{{\bf j}} \langle
a^{\phantom{\dagger}}_{{\bf j}}(\tau) a^{\dagger}_{{\bf
j}\pm}(0) \rangle$ via
\begin{equation}
   G^{a}(\tau)= \int_{-\infty}^{\infty}\ d\omega \frac{e^{-\omega\tau}}{e^{-\beta\omega}+1}\ N^{a}(\omega) \label{aw}
\end{equation}
To avoid the ambiguity from analytical continuation such as maximum entropy method\cite{gubernatis91}, we resort to the approximate formula $N^a(\omega=0) \approx \beta G^a(\tau=\beta/2)/\pi  \label{Gtau}$
assuming that the temperature is much lower than the energy scale on which there are structures in DOS\cite{GtauNw}.

As shown in Figure~\ref{fig4}, the dominant feature associated with both the $f_2$-AF$^M$ and $c$-$f_2$-M phases at large $t_{\perp}\geq V$ is the metallic behavior of both $c$ and $f_2$ orbitals while $f_1$ orbital is readily insulating. 
The stronger metallicity of the conduction electron can be easily understood as the consequence of significantly weakened $c$-$f_1$ spin correlation (Fig.~\ref{fig3}) that liberates the $c$ electrons so that $\beta G^{c}(\beta/2)$ keeps growing with $t_{\perp}$. Strikingly, the comparison with Fig.~\ref{saf} demonstrates that the metallicity of $f_2$ starts within the phase with $f_2$-AF order (cyan regime in Fig.~\ref{phase}). Therefore, the $f_2$-AF$^M$ phase displays the coexistence and competition of the metallicity and AF order of $f_2$ orbital. In fact, the metallicity participates in destroying the $f_2$-AF order.
The distinct difference between $f_1$ and $f_2$ reflects the more freedom of $f_2$ despite of the gradually stronger $f_1$-$f_2$ binding, which is consistent with the smoother variation of $f_1$ local moment with $t_{\perp}$ in Fig.~\ref{fig3}(b). Apparently, we confirm that $\beta G^{f_2}(\beta/2)$ finally vanishes at sufficiently large $t_{\perp}$, e.g. at $V/t=1.2$ (blue solid line), where the system becomes strongly bound $f_1$-$f_2$ dimers plus nearly free conduction electrons.

\section{Conclusion}
In conclusion, as a proof of principle study, we have addressed the fundamental question of whether or not the antiferromagnetic (AF) order can be induced via the local moments' hybridization with the heavy electrons instead of conduction electrons. We provided strong numerical evidence to confirm its possibility via a prototypical model through determinant QMC simulations. In particular, we claim that this AF order mediated by heavy electrons is realized by a so-called ``high-order'' RKKY interaction that resembles the conventional RKKY mediated via the conduction electrons in standard PAM/KLM. We emphasize that the induced AF order only emerges if the heavy electrons are present, whose breakdown coincides with the disappearance of the ordering.
Moreover, we further prove that the induced AF order can coexist with its metallicity in a finite regime of the phase diagram, which competes with and ultimately destroys the AF order. 

As our motivation partly came from the potential relevance to the heavy fermion compound Ce$_3$(Pt/Pd)In$_{11}$~\cite{Custers2015,Custers15,Kaczorowski18,Custers2020}, we remark that the three orbitals $c,f_1,f_2$ in our prototypical model can be used to mimic Pt/Pd, Ce(1), and Ce(2) separately of Ce$_3$(Pt/Pd)In$_{11}$~\cite{private}. Our findings implies that the experimentally observed magnetic ordering of Ce(2) ($f_2$ orbital) can indeed coexist microscopically with the fully Kondo screened Ce(1) ($f_1$ orbital) and in fact the Ce(1) plays a significant role in forming the AF order of Ce(2) sublattice. 
To some extent, however, our model has intrinsic limitation due to its neglecting of conduction electron reservoir from In sites because it has been shown that the strong hybridization with the out of plane In plays an important role in other Ce-based compounds, such as Ce-115 materials~\cite{Ce115}. 
Therefore, it is requisite to explore the more appropriate models for the potential connection of our findings reported here to the realistic materials, which is left for future investigation. Another fascinating theoretical question regards on the reverse role of Ce(2) on the superconductivity claimed experimentally to be responsible by Ce(1)~\cite{Custers2020}.
Besides, the thorough understanding and realization of the proposed ``high-order'' RKKY interaction in other contexts would be highly interesting.

\begin{acknowledgments}
We acknowledge Richard Scalettar and Jeroen Custers for fruitful discussion in the initial stage. This work was funded by the Stewart Blusson Quantum Matter Institute at University of British Columbia, and by the Natural Sciences and Engineering Research Council of Canada.
\end{acknowledgments}


\end{document}